\documentclass[conference]{IEEEtran}
\usepackage{amsmath,amssymb,amsfonts,dsfont,verbatim,cite,amsopn,graphicx,citesort,url,color,subfigure,enumitem}

\newcommand{\Pmax}{P_\mathrm{max}}
\newcommand{\Pc}{\mathbb{P}_\mathrm{cov}}
\newcommand{\alphab}{\alpha_\mathrm{TBS}}
\newcommand{\alphacd}{\alpha_\mathrm{TA}}
\newcommand{\alphadd}{\alpha_\mathrm{AA}}
\newcommand{\rhob}{\rho_\mathrm{T}}
\newcommand{\rhod}{\rho_\mathrm{A}}
\newcommand{\gammab}{\gamma_\mathrm{T}}
\newcommand{\gammad}{\gamma_\mathrm{A}}
\newcommand{\Ib}{I_\mathrm{T}}
\newcommand{\Id}{I_\mathrm{A}}
\newcommand{\Zd}{\mathcal{Z}_A}
\newcommand{\Zc}{\mathcal{Z}_T}
\newcommand{\mcd}{m_\mathrm{TA}}
\newcommand{\mdd}{m_\mathrm{AA}}

\newtheorem{theorem}{Theorem}
\newtheorem{lemma}{Lemma}

\begin{document}
\title{Uplink Coverage Performance of an Underlay Drone Cell for Temporary Events}
\author{\IEEEauthorblockN{Xiaohui Zhou${}^\ast$, Jing Guo${}^\ast$, Salman Durrani${}^\ast$, and Halim Yanikomeroglu${}^\dag$}
\IEEEauthorblockA{${}^\ast$Research School of Engineering, The Australian National University, Canberra, ACT 2601, Australia.\\
Emails: \{xiaohui.zhou, jing.guo, salman.durrani\}@anu.edu.au.\\
${}^\dag$Department of Systems and Computer Engineering, Carleton University, Ottawa, ON K1S 5B6, Canada.\\
Email: halim@sce.carleton.ca.}}
\IEEEspecialpapernotice{\Large (Invited Paper)}

\maketitle

\begin{abstract}
Using a drone as an aerial base station (ABS) to provide coverage to users on the ground is envisaged as a promising solution for beyond fifth generation (beyond-5G) wireless networks. While the literature to date has examined downlink cellular networks with ABSs, we consider an uplink cellular network with an ABS. Specifically, we analyze the use of an underlay ABS to provide coverage for a temporary event, such as a sporting event or a concert in a stadium. Using stochastic geometry, we derive the analytical expressions for the uplink coverage probability of the terrestrial base station (TBS) and the ABS. The results are expressed in terms of (i) the Laplace transforms of the interference power distribution at the TBS and the ABS and (ii) the distance distribution between the ABS and an independently and uniformly distributed (i.u.d.) ABS-supported user equipment and between the ABS and an i.u.d. TBS-supported user equipment. The accuracy of the analytical results is verified by Monte Carlo simulations. Our results show that varying the ABS height leads to a trade-off between the uplink coverage probability of the TBS and the ABS. In addition, assuming a quality of service of 90\% at the TBS, an uplink coverage probability of the ABS of over 85\% can be achieved, with the ABS deployed at or below its optimal height of typically between $250-500$ m for the considered setup.
\end{abstract}

\vspace{-2mm}
\section{Introduction}\label{sec:Intro}
The use of drones as aerial base stations (ABSs), to form drone cells and to provide flexible and agile coverage, has gained enormous interest as a potential beyond fifth generation (beyond-5G) wireless network solution~\cite{Yaliniz-2016,Chandrasekharan-2016}. In \cite{Yaliniz-2016}, eight scenarios have been identified for drone cell deployment, with drones providing service to: 1) rural areas, 2) urban areas, 3) users with high mobility, 4) congested urban areas, 5) congested backhaul, 6) temporary events, 7) temporary blind spots and 8) sensor networks. For these scenarios, drone cells have been investigated in the literature from different perspectives, such as drone channel modeling~\cite{Amorim-2017a,Amorim-2017b,Hourani-2018,Hourani-2014}, drone deployment and optimization of trajectories~\cite{Fotouhi-2017,Zeng-2016a,Alzenad-2017a,He-2018,Zhang-2017,Lagum-2017,Wu-2017,Lyu-2017}, and performance analysis of drone enabled cellular systems~\cite{Mozaffari-2016c,Chetlur-2017,Azari-2017a,Galkin-2017,Azari-2017b}. In this work, we focus on the last aspect.{\let\thefootnote\relax\footnotetext{This work was supported in part by the Australian Research Council's Discovery Project Funding Scheme (Project number  DP170100939) and the Natural Sciences and Engineering Research Council of Canada (NSERC).}}

\textit{\underline{Motivation:}} The literature to date~\cite{Mozaffari-2016c,Chetlur-2017,Azari-2017a,Galkin-2017,Azari-2017b} has focused on the downlink, generally for Scenarios 1--3. In this work, we focus on the uplink for Scenario 6, i.e., using ABS to provide additional coverage for temporary events, such as concerts or sports events. Such temporary events are happening more frequently in cities all over the world with very high number of users gathering, which often leads to network congestion and poor user experience. The optimal ABS height with respect to the downlink performance does not necessarily optimize the uplink performance. For example, the distribution of the interference powers are different for the uplink and downlink cases. Moreover, the uplink is the bottleneck in such events since an increasing number of users try to share content from the event on social media applications. Therefore, assessing the usefulness of ABSs to provide uplink coverage for temporary events is an important open problem in the literature which has great practical importance as well.

\textit{\underline{Related Work:}} The downlink performance of a single static drone and a single mobile drone with underlay device-to-device users was studied in \cite{Mozaffari-2016c}, while \cite{Chetlur-2017} analyzed the downlink coverage of a finite network formed by multiple drones. The downlink coverage probability of a network with multiple directional beamforming drones was investigated in \cite{Azari-2017a}. The downlink coverage performance of a network of multiple drones in an urban environment was studied in\cite{Galkin-2017}. In \cite{Azari-2017b}, the authors studied the downlink in a cellular network with multiple ground base stations and a drone user equipment (UE).

\textit{\underline{Contributions:}} In this paper, we analyze the use of an ABS to provide coverage for a temporary event. To the best of our knowledge, this scenario has not yet been considered in the literature to date. Using stochastic geometry, we analyze the uplink coverage probability of the terrestrial base station (TBS) and the ABS. The novel contributions of this paper are:
\begin{itemize}
\item We derive the analytical expressions for uplink coverage probability of the TBS and the ABS. The results are in terms of (i) the Laplace transforms of the interference power distribution at the TBS and the ABS and (ii) the distance distribution between the ABS and an independently and uniformly distributed (i.u.d.) ABS-supported user equipment (AsUE) and between the ABS and an i.u.d. TBS-supported user equipment (TsUE).
\item Our results show that increasing the height of the ABS can generally improve the uplink coverage probability of the ABS, while it degrades the uplink coverage probability of the TBS. In addition, there is an optimal ABS height which maximizes the uplink coverage probability of the ABS.
\item We assess the feasibility of establishing an underlay ABS to provide uplink coverage for temporary events. Assuming a quality of service of 90\% at the TBS, under our considered system set up, uplink coverage probability of the ABS of over 85\% can be achieved, with the ABS at or below its optimal height if the center of the stadium is sufficiently distanced from the TBS.
\end{itemize}

\vspace{-2mm}
\section{System model}\label{sec:systemmodel}
\begin{figure}[t]
\centering
\subfigure[3D view.]{\includegraphics[width=0.24\textwidth]{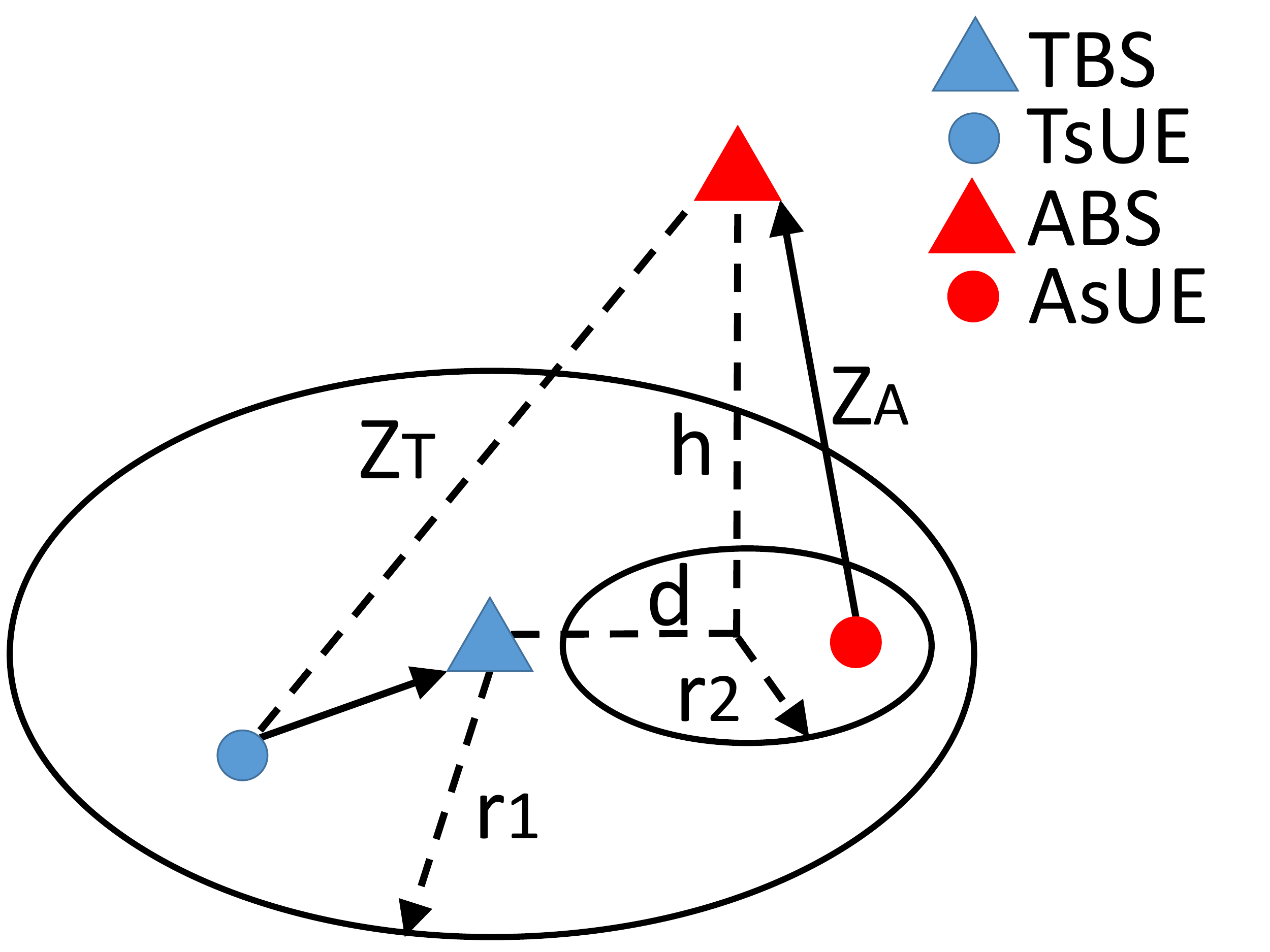}}
\subfigure[Projection on the ground.]{\includegraphics[width=0.24\textwidth]{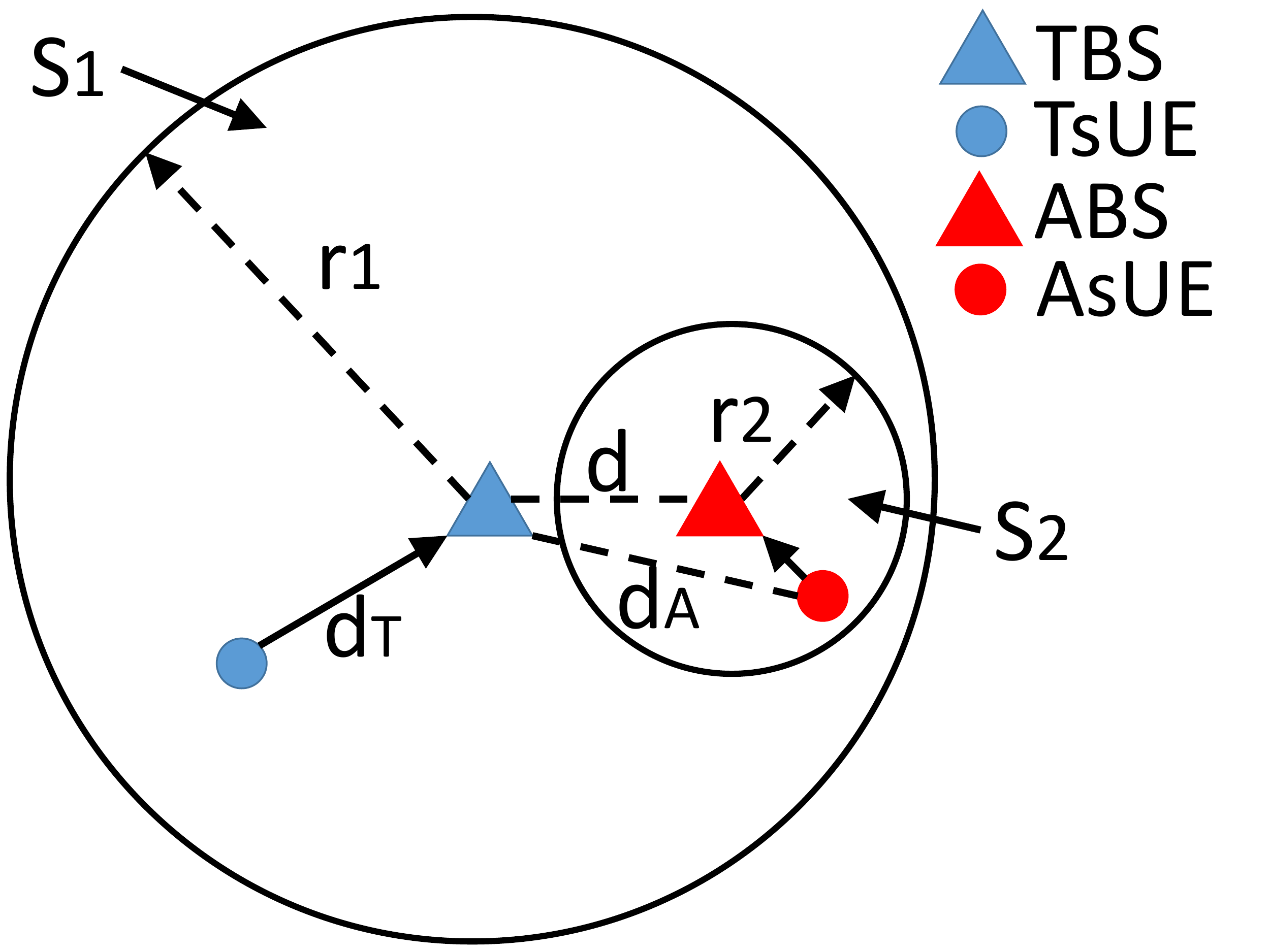}}
\vspace{-3mm}
\caption{Illustration of the system model.}
\label{fig:system}
\vspace{-6mm}
\end{figure}

\vspace{-0.2mm}
We consider the uplink communication in a two-cell network comprised of a TBS and an ABS, where the network region $\mathcal{S}_1$ is a disk with radius $r_1$, i.e., $|\mathcal{S}_1|=\pi r_1^2$ and a TBS is located at the center. We assume that there is a temporary event held inside a stadium within the network region. The stadium's building area $\mathcal{S}_2$ is modeled as a disk with radius $r_2$ and its center is at a distance $d$ from the TBS. To provide additional resources for the event, an ABS is deployed to act as an ABS\footnote{Current drone regulations prohibit a drone from flying over stadiums or sports events. This is expected to change in the future.}. The ABS is assumed to be placed at a height of $h$ above the center of the stadium, as shown in Fig.~\ref{fig:system}. There are $N_c$ TsUEs served by the TBS, which are spatially distributed according to a Binomial point process (BPP) inside the network region excluding the stadium, i.e., $\mathcal{S}_1\setminus\mathcal{S}_2$. At the same time, there are $N_d$ AsUEs on the ground inside the stadium $\mathcal{S}_2$. For tractability, we assume that all AsUEs are served by the ABS only. The location of AsUEs is modelled as an independent BPP in $\mathcal{S}_2$.

\textit{\underline{Channel Model:}} There are two types of communication links in the considered system model: aerial links and terrestrial links. The link between the TsUE and the ABS and the link between the AsUE and the ABS are aerial links. The link between the TsUE and the TBS and the link between the AsUE and the TBS are terrestrial links. For analytical tractability, similar to \cite{Chetlur-2017,Azari-2017b}, we assume the aerial links experience Nakagami-$m$ fading. The AsUE to ABS link has a strong line of sight (LOS) with fading parameter $\mdd$ while the TsUE to ABS link has a weak LOS with fading parameter $\mcd$. As for the terrestrial links, Rayleigh fading is assumed~\cite{Zhang-2017}.
%\footnote{The prior works on the literature~\cite{Amorim-2017a,Amorim-2017b,Hourani-2017,Hourani-2014} focus on the air-to-ground and cellular-to-UAV propagation modeling for downlink communications. Other uplink channel models for aerial link could be considered for extension work.}

A general power-law path-loss model is considered\footnote{The validity of using this path-loss model for aerial links will be verified in Section~\ref{sec:result} by comparing with simulations using air-to-ground channel model for aerial links.}, in which the signal power decays at a rate $\ell^{-\alpha}$ with the propagation distance $\ell$ and $\alpha$ is the path-loss exponent. Due to different characteristics of aerial link and terrestrial link, different path-loss exponent $\alphab$, $\alphacd$ and $\alphadd$ are assumed for terrestrial link, aerial link between the TsUE and the ABS, and aerial link between the AsUE and the ABS respectively. Furthermore, both the TsUE to TBS link and the AsUE to ABS link experience additive white Gaussian noise (AWGN) with variance $\sigma^2$.

\textit{\underline{Uplink Network Model:}} For the spectrum efficiency, we assume that both the TBS and the ABS share the same spectrum resource, i.e., in an underlay fashion. Orthogonal multiple access technique is employed in this work. We assume that the number of the TsUEs and the AsUEs are sufficiently high. That is to say, there will always be one TsUE and one AsUE to be served per each channel at the same time. Hence, there is no interference among TsUEs (or AsUEs), but interference exists between TsUEs and AsUEs. In this work, we focus our analysis on one uplink channel since other channels share the same interference statistics.

For reliable and successful uplink communication, the TsUE controls its transmit power such that the average signal received at the TBS is equal to the threshold $\rhob$. Power control is deployed at the AsUE. We also set a maximum transmit power constraint at the AsUE, to avoid the transmit power for AsUE going to very large when the ABS is placed at a high altitude. In other words, the AsUE compensates for the path-loss to keep the average signal power at the ABS equal to the threshold $\rhod$ if the transmit power required for the path-loss inversion is less than $\Pmax$. Otherwise, the AsUE tries to establish an uplink connection with the ABS by transmitting with a power of $\Pmax$. Therefore, the instantaneous transmit power for the AsUE, $P_a$, depends on the propagation distance between the AsUE and the ABS and can be shown as \eqref{eq:Pu} at the top of the next page, where $\Zd$ is the Euclidean distance between the AsUE and the ABS.

\begin{figure*}[t]
\begin{align}\label{eq:Pu}
P_a=\begin{cases}
                \Pmax,& h\geqslant(\frac{\Pmax}{\rhod})^{\frac{1}{\alphadd}}\;\;\; ||\;\;\; \left(\sqrt{(\frac{\Pmax}{\rhod})^{\frac{2}{\alphadd}}-r_2^2}<h<(\frac{\Pmax}{\rhod})^{\frac{1}{\alphadd}}\;\;\; \&\;\;\; \Zd>(\frac{\Pmax}{\rhod})^{\frac{1}{\alphadd}}\right)\\
				\rhod \Zd^{\alphadd},& h\leqslant\sqrt{(\frac{\Pmax}{\rhod})^{\frac{2}{\alphadd}}-r_2^2}\;\;\; || \;\;\; \left(\sqrt{(\frac{\Pmax}{\rhod})^{\frac{2}{\alphadd}}-r_2^2}<h<(\frac{\Pmax}{\rhod})^{\frac{1}{\alphadd}}\;\;\; \& \;\;\;\Zd<(\frac{\Pmax}{\rhod})^{\frac{1}{\alphadd}}\right)
		\end{cases}.
\end{align}
\rule{18.2cm}{0.5pt}
\vspace{-10mm}
\end{figure*}

\textit{\underline{SINR:}} For the considered setup, the instantaneous signal-to-interference-plus-noise ratio (SINR) at the TBS is given as
\begin{align}\label{eq:sinrBS}
\textsf{SINR}_\mathrm{T}\!=\!\frac{P_t H_T d_T^{-\alphab}}{P_a H_A d_A^{-\alphab}+\sigma^2}\!=\!\frac{\rhob H_T}{P_a H_A d_A^{-\alphab}+\sigma^2},
\end{align}
where $P_t=\rhob d_T^{\alphab}$ is the TsUE transmit power. $H_T$ and $H_A$ are the fading power gain between the TsUE and the TBS and between the AsUE and the TBS, respectively, which follow exponential distribution. $d_T$ and $d_A$ are the Euclidean distance between the TsUE and the TBS and between the AsUE and the TBS, respectively. The transmit power of the AsUE $P_a$ is given in \eqref{eq:Pu}.

The instantaneous SINR at the ABS is given as
\begin{align}\label{eq:sinrUAV}
\textsf{SINR}_\mathrm{A}=\frac{P_a G_A \Zd^{-\alphadd}}{P_t G_T \Zc^{-\alphacd}+\sigma^2},
\end{align}
where $G_A$ and $G_T$ are the fading power gain between the AsUE and the ABS and between the TsUE and the ABS, respectively, which follow Gamma distribution. $\Zd$ and $\Zc$ are the Euclidean distance between the AsUE and the ABS and between the TsUE and the ABS, respectively.

\setcounter{equation}{4}
\begin{figure*}[t]
\begin{align}\label{eq:LIB}
\mathcal{L}_{\Ib}(s)=\begin{cases}
                \int_h^{\sqrt{h^2+r_2^2}}\int_{0}^{2\pi}\mathcal{L}_{\Ib}(s,\Pmax|\theta,z)\frac{1}{2\pi}f_{\Zd}(z)\textup{d}\theta \textup{d}z, &h\geqslant(\frac{\Pmax}{\rhod})^{\frac{1}{\alphadd}}\\
                \int_h^{(\frac{\Pmax}{\rhod})^\frac{1}{\alphadd}}\int_{0}^{2\pi}\mathcal{L}_{\Ib}(s,\rhod z^{\alphadd}|\theta,z)\frac{1}{2\pi}f_{\Zd}(z)\textup{d}\theta \textup{d}z\\
                +\int_{(\frac{\Pmax}{\rhod})^\frac{1}{\alphadd}}^{\sqrt{h^2+r_2^2}}\int_{0}^{2\pi}\mathcal{L}_{\Ib}(s,\Pmax|\theta,z)\frac{1}{2\pi}f_{\Zd}(z)\textup{d}\theta \textup{d}z, &\sqrt{(\frac{\Pmax}{\rhod})^{\frac{2}{\alphadd}}-r_2^2}<h<(\frac{\Pmax}{\rhod})^{\frac{1}{\alphadd}}\\
                \int_h^{\sqrt{h^2+r_2^2}}\int_{0}^{2\pi}\mathcal{L}_{\Ib}(s,\rhod z^{\alphadd}|\theta,z)\frac{1}{2\pi}f_{\Zd}(z)\textup{d}\theta \textup{d}z, &h\leqslant\sqrt{(\frac{\Pmax}{\rhod})^{\frac{2}{\alphadd}}-r_2^2}\\
		\end{cases}.
\end{align}
\rule{18.2cm}{0.5pt}
\vspace{-10mm}
\end{figure*}

\vspace{-2mm}
\section{Coverage Probability}\label{sec:analysis}
In this section, we analyze the network performance by adopting coverage probability as the performance metric. The coverage probability is formally defined as
\setcounter{equation}{3}
\begin{align}
\Pc^b\triangleq\Pr(\textsf{SINR}_b>\gamma_b),
\end{align}
where superscript $b$ is $\mathrm{T}$ for TBS and $\mathrm{A}$ for ABS, and $\gamma_b$ is the SINR threshold. $\textsf{SINR}_\mathrm{T}$ and $\textsf{SINR}_\mathrm{A}$ can be found in \eqref{eq:sinrBS} and \eqref{eq:sinrUAV}, respectively. The results for the coverage probability of the TBS and the ABS are presented in the next two subsections.

\vspace{-2mm}
\subsection{TBS Coverage Probability}
First we present two lemmas, which are used in deriving the coverage probability of the TBS in Theorem~\ref{th:covBS}.

\begin{lemma}\label{le:LIB}
The Laplace transform of the interference power distribution at the TBS is given as \eqref{eq:LIB} at the top of the this page,
where
\setcounter{equation}{5}
\begin{align}
\mathcal{L}_{\Ib}(s,p|\theta,z)\!=\!\frac{1}{1\!+\!\frac{sp}{\left(z^2\!-\!h^2\!+\!d^2\!-\!2\sqrt{z^2\!-\!h^2}d\cos(\theta)\right)^{\frac{\alphab}{2}}}}.
\end{align}
\end{lemma}
\begin{IEEEproof}
See Appendix~\ref{ap:LIB}.
\end{IEEEproof}

From Lemma~\ref{le:LIB}, we can see that the transmit power of the AsUE $P_a$ and the distance between the AsUE and the TBS $d_A$ are related to the distance between the AsUE and the ABS $\Zd$. This important distance distribution is presented in the following lemma.
\begin{lemma}\label{le:disUU}
The probability density function (pdf) of the distance $\Zd$ between the ABS at height $h$ above the center of $\mathcal{S}_2$ and an i.u.d. AsUE inside $\mathcal{S}_2$ is
\begin{align}\label{eq:fZd}
f_{\Zd}(z)=\frac{2z}{r_2^2}, \;\;h\leqslant z\leqslant\sqrt{r_2^2+h^2}.
\end{align}
\end{lemma}
\begin{IEEEproof}
See Appendix~\ref{ap:disUU}.
\end{IEEEproof}

\begin{theorem}\label{th:covBS}
Based on the system model in Section~\ref{sec:systemmodel}, the coverage probability of the TBS is
\begin{align}\label{eq:covB}
\Pc^\mathrm{T}=\exp\left(-\frac{\gammab}{\rhob}\sigma^2\right)\mathcal{L}_{\Ib}(s),
\end{align}
where $\Ib=P_a H_A d_A^{-\alphab}$, $s=\frac{\gammab}{\rhob}$, and $\mathcal{L}_{\Ib}(s)$ is given by Lemma~\ref{le:LIB}.
\end{theorem}
\begin{IEEEproof}
See Appendix~\ref{ap:covBS}.
\end{IEEEproof}

Substituting \eqref{eq:LIB} and \eqref{eq:fZd} into \eqref{eq:covB}, we can obtain the coverage probability of the TBS.

\vspace{-2mm}
\subsection{ABS Coverage Probability}
We begin by presenting three lemmas, which will then be used to compute the coverage probability of the ABS in Theorem~\ref{th:covD}.

\begin{lemma}\label{le:LIU}
The Laplace transform of the interference power distribution at the ABS is
\begin{align}
&\mathcal{L}_{\Id}\!(s)\!=\!\int_{\sqrt{r_2^2+h^2}}^{\sqrt{(r_1-d)^2+h^2}}\!\!\int_{0}^{2\pi}\!\!\!\!\mathcal{L}_{\Id}(s|\omega,z)f_\Omega(\omega|z)f_{\Zc}(z)\textup{d}\omega \textup{d}z\nonumber\\
&+\int_{\sqrt{(r_1-d)^2+h^2}}^{\sqrt{(r_1+d)^2+h^2}}\int_{-\widehat{\omega}}^{\widehat{\omega}}\mathcal{L}_{\Id}(s|\omega,z)f_\Omega(\omega|z)f_{\Zc}(z)\textup{d}\omega \textup{d}z,
\end{align}
where
\begin{align}
&\mathcal{L}_{\Id}(s|\omega,z)=\mcd^{\mcd}\left(\mcd+s\rhob z^{-\alphacd}\right.\nonumber\\
&\left.\times \left(z^2\!-\!h^2\!+\!d^2\!-\!2\sqrt{z^2\!-\!h^2}d\cos(\omega)\right)^{\frac{\alphab}{2}}\right)^{-\mcd}.
\end{align}
%and $\widehat{\omega}=\mathrm{arcsec}\left(\frac{2d\sqrt{z^2-h^2}}{d^2+z^2-h^2-r_1^2}\right)$.
\end{lemma}
\begin{IEEEproof}
The proof follows the same lines as Lemma~\ref{le:LIB} and is skipped for the sake of brevity.
\end{IEEEproof}

The pdf of the distance between the TsUE and the ABS $f_{\Zc}(z)$, and the conditional pdf of the angle, $f_\Omega(\omega|z)$, between the ground projection of $\Zc$ and $d_T$ are given in Lemma~\ref{le:disCU} and Lemma~\ref{le:degCU}, respectively.

\begin{lemma}\label{le:disCU}
The pdf of the distance $\Zc$ between the ABS at height $h$ above the center of $\mathcal{S}_2$ and an i.u.d. TsUE inside $\mathcal{S}_1\setminus\mathcal{S}_2$ is
\begin{align}\label{eq:disCU}
f_{\Zc}\!(\!z\!)\!=\!\!\begin{cases}\frac{2z}{r_1^2\!-\!r_2^2}, &\sqrt{r_2^2\!+\!h^2}\!\leqslant\! z\!\leqslant\!\sqrt{(r_1\!-\!d)^2\!+\!h^2}\\
            \frac{2z\widehat{\omega}}{\pi (r_1^2\!-\!r_2^2)}, &\sqrt{(r_1\!-\!d)^2\!+\!h^2}\!\leqslant\! z\!\leqslant\!\sqrt{(r_1\!+\!d)^2\!+\!h^2}
\end{cases},
\end{align}
where $\widehat{\omega}=\mathrm{arcsec}(\frac{2d\sqrt{z^2-h^2}}{d^2+z^2-h^2-r_1^2})$.
\end{lemma}
\begin{IEEEproof}
See Appendix~\ref{ap:disCU}.
\end{IEEEproof}

\begin{lemma}\label{le:degCU}
The pdf of the angle, $f_\Omega(\omega|z)$, between the ground projection of $\Zc$ and $d_T$ conditioned on $\Zc$ is
\begin{align}\label{eq:degCU}
f_\Omega(\omega|z)\!=\!\begin{cases}\frac{1}{2\pi}, &\sqrt{r_2^2\!+\!h^2}\!\leqslant\! z\!\leqslant\!\sqrt{(r_1\!-\!d)^2\!+\!h^2}\\
            \frac{1}{2\widehat{\omega}}, &\sqrt{(r_1\!-\!d)^2\!+\!h^2}\!\leqslant\! z\!\leqslant\!\sqrt{(r_1\!+\!d)^2\!+\!h^2}
\end{cases}.
\end{align}
\end{lemma}
\begin{IEEEproof}
This lemma can be proved by using cosine rule and simple trigonometry.
\end{IEEEproof}

\begin{figure*}[t]
\begin{align}\label{eq:covD}
\Pc^\mathrm{A}\!\!=\!\!\begin{cases}
                \int_h^{\sqrt{h^2\!+r_2^2}}\!\Pc^\mathrm{A}(\Pmax|z)f_{\Zd}(z)\textup{d}z,& h\geqslant(\frac{\Pmax}{\rhod})^{\frac{1}{\alphadd}}\\
                \int_h^{(\frac{\Pmax}{\rhod})^{\frac{1}{\alphadd}}}\!\Pc^\mathrm{A}(\rhod z^{\alphadd}|z)f_{\Zd}(z)\textup{d}z\\
								+\!\int_{(\frac{\Pmax}{\rhod})^{\frac{1}{\alphadd}}}^{\sqrt{h^2\!+r_2^2}}\!\Pc^\mathrm{A}(\Pmax|z)f_{\Zd}(z)\textup{d}z,& \sqrt{(\frac{\Pmax}{\rhod})^{\frac{2}{\alphadd}}\!-\!r_2^2}\!<\!h\!<\!(\frac{\Pmax}{\rhod})^{\frac{1}{\alphadd}}\\
                \int_h^{\sqrt{h^2\!+r_2^2}}\!\Pc^\mathrm{A}(\rhod z^{\alphadd}|z)f_{\Zd}(z)\textup{d}z,& h\leqslant\sqrt{(\frac{\Pmax}{\rhod})^{\frac{2}{\alphadd}}-r_2^2}\\
		\end{cases}.
\end{align}
\rule{18.2cm}{0.5pt}
\vspace{-8mm}
\end{figure*}
\begin{figure*}[t]
\begin{minipage}[t]{0.45\linewidth}
\centering
\includegraphics[width=0.9 \textwidth]{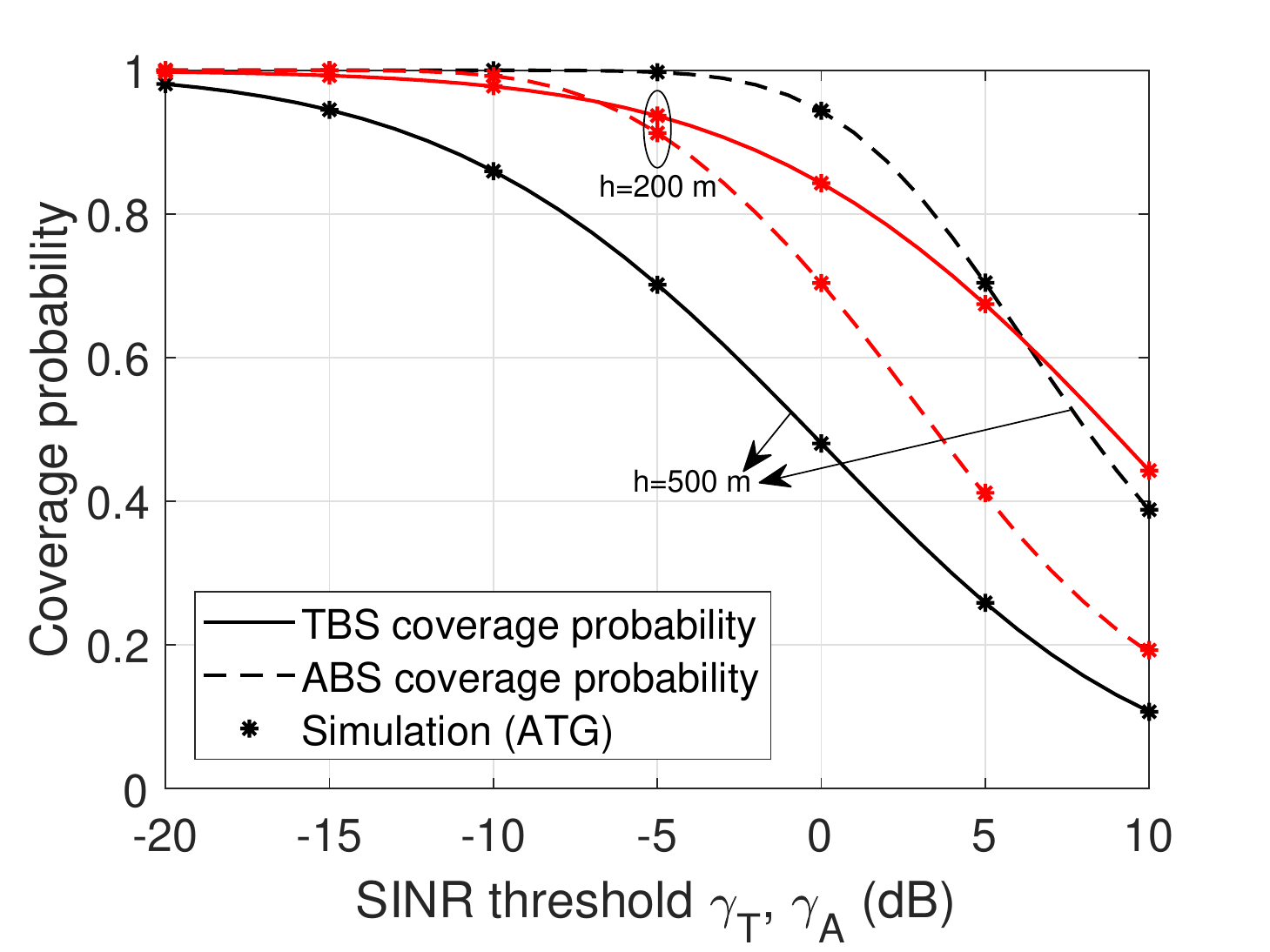}
\centering
\vspace{-2mm}
\caption{Coverage probability of the TBS and the ABS versus the SINR threshold $\gammab$ and $\gammad$ with different height of the ABS and simulation with ATG aerial link model.}
\centering
\label{fig:gamma}
\end{minipage}\hfill
\centering
\begin{minipage}[t]{0.45\linewidth}
\centering
\includegraphics[width=0.9 \textwidth]{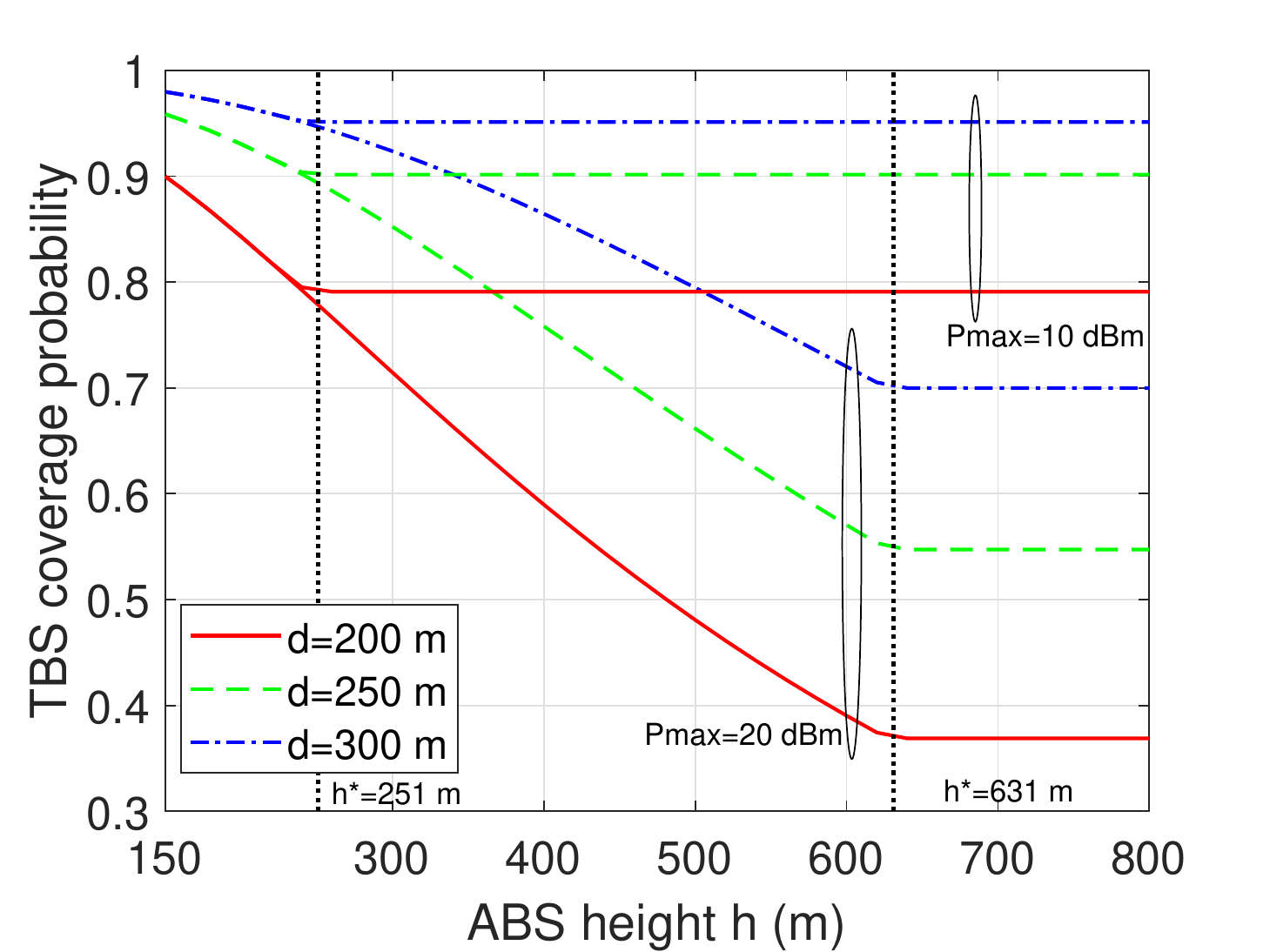}
\centering
\vspace{-2mm}
\caption{Coverage probability of the TBS versus the height of the ABS with different distance between the center of the stadium and the TBS and different maximum transmit power for the AsUE.}
\centering
\label{fig:h_bs}
\end{minipage}
\vspace{-6mm}
\end{figure*}

\begin{theorem}\label{th:covD}
Based on the system model in Section~\ref{sec:systemmodel}, the coverage probability of the ABS is given as \eqref{eq:covD} at the top of this page, where
\begin{align}
\Pc^\mathrm{A}(P_a|z)=&\sum_{n=0}^{\mdd-1}\frac{(-s)^n}{n!}\exp(-s\sigma^2)\nonumber\\
&\times\sum_{k=0}^n\binom{n}{k}(-\sigma^2)^{n-k}\frac{\textup{d}^k}{\textup{d}s^k}\mathcal{L}_{\Id}(s),
\end{align}
$s=\frac{\mdd\gammad z^{\alphadd}}{p}$ and $\mathcal{L}_{\Id}(s)$ is given by Lemma~\ref{le:LIU}. The pdf of the distance between the AsUE and the ABS $f_{\Zd}(z)$ is provided in Lemma~\ref{le:disUU}.
\end{theorem}
\begin{IEEEproof}
See Appendix~\ref{ap:covD}.
\end{IEEEproof}

Combining Lemma~\ref{le:LIU},~\ref{le:disCU}, and~\ref{le:degCU} with Theorem~\ref{th:covD}, we can calculate the coverage probability of the ABS.

\vspace{-2mm}
\section{Results}\label{sec:result}
In this section, we first validate the analytical results and then discuss the design insights. The simulation results are generated by averaging over $10^6$ Monte Carlo simulation runs. In the simulations, the terrestrial links follow the power-loss path-loss model described in Section~\ref{sec:systemmodel}. However, the aerial links follow the air-to-ground (ATG) channel model in~\cite{Mozaffari-2016c} with $C=4.88$, $B=0.43$ and $\eta=-20$ dB. Unless stated otherwise, we set the parameters as follows: $r_1=500$ m, $r_2=100$ m, $d=200$ m, $\Pmax=20$ dBm, $\rhob=-75$ dBm, $\rhod=-50$ dBm, $\alphadd=2.5$, $\alphacd=3$, $\alphab=4$, $\mdd=5$, $\mcd=3$, $\sigma^2=-100$ dBm, $\gammad=0$ dB and $\gammab=0$ dB.

Fig.~\ref{fig:gamma} plots the coverage probability of the TBS and the ABS against the SINR threshold with an ABS height of $200$ m and $500$ m. The simulation results match very well with the analytical results. This is because that the ABS is assumed to be placed directly above the stadium in our model and the likelihood of the aerial links being LOS is very high. This validates the use of the tractable power-law path-loss model for aerial links in our analysis. From this figure, we can see that if an ABS is placed at a higher altitude, the coverage probability of the ABS is higher, but the coverage probability of the TBS is lower. This is discussed in detail next.
\begin{figure*}[t]
\begin{minipage}[t]{0.45\linewidth}
\centering
\includegraphics[width=0.9 \textwidth]{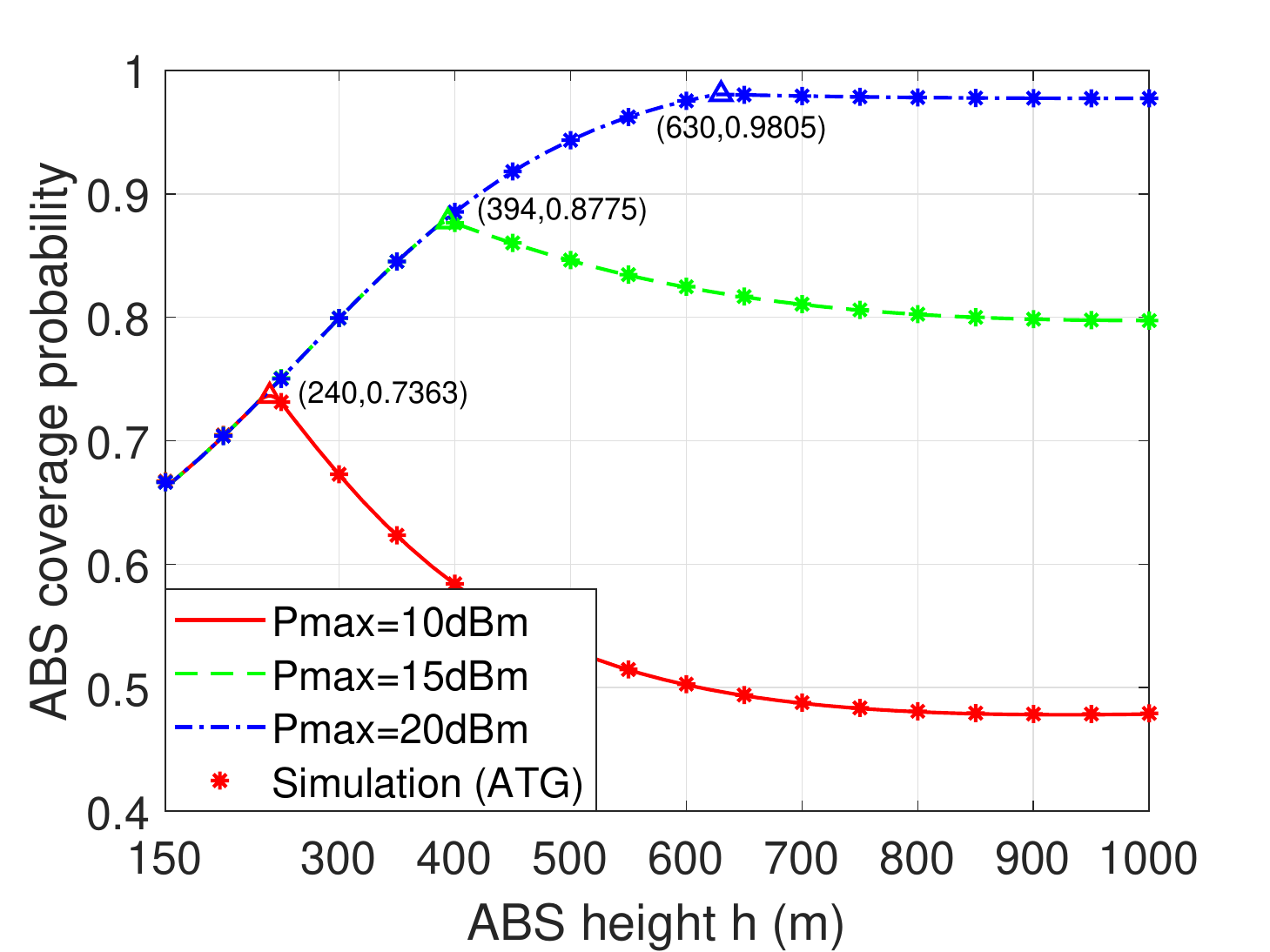}
\centering
\vspace{-2mm}
\caption{Coverage probability of the ABS versus the ABS height with different AsUE maximum transmit power and simulation with ATG aerial link model. The triangle markers denote the optimal ABS height.}
\centering
\label{fig:h_uav}
\end{minipage}\hfill
\centering
\begin{minipage}[t]{0.45\linewidth}
\centering
\includegraphics[width=0.9 \textwidth]{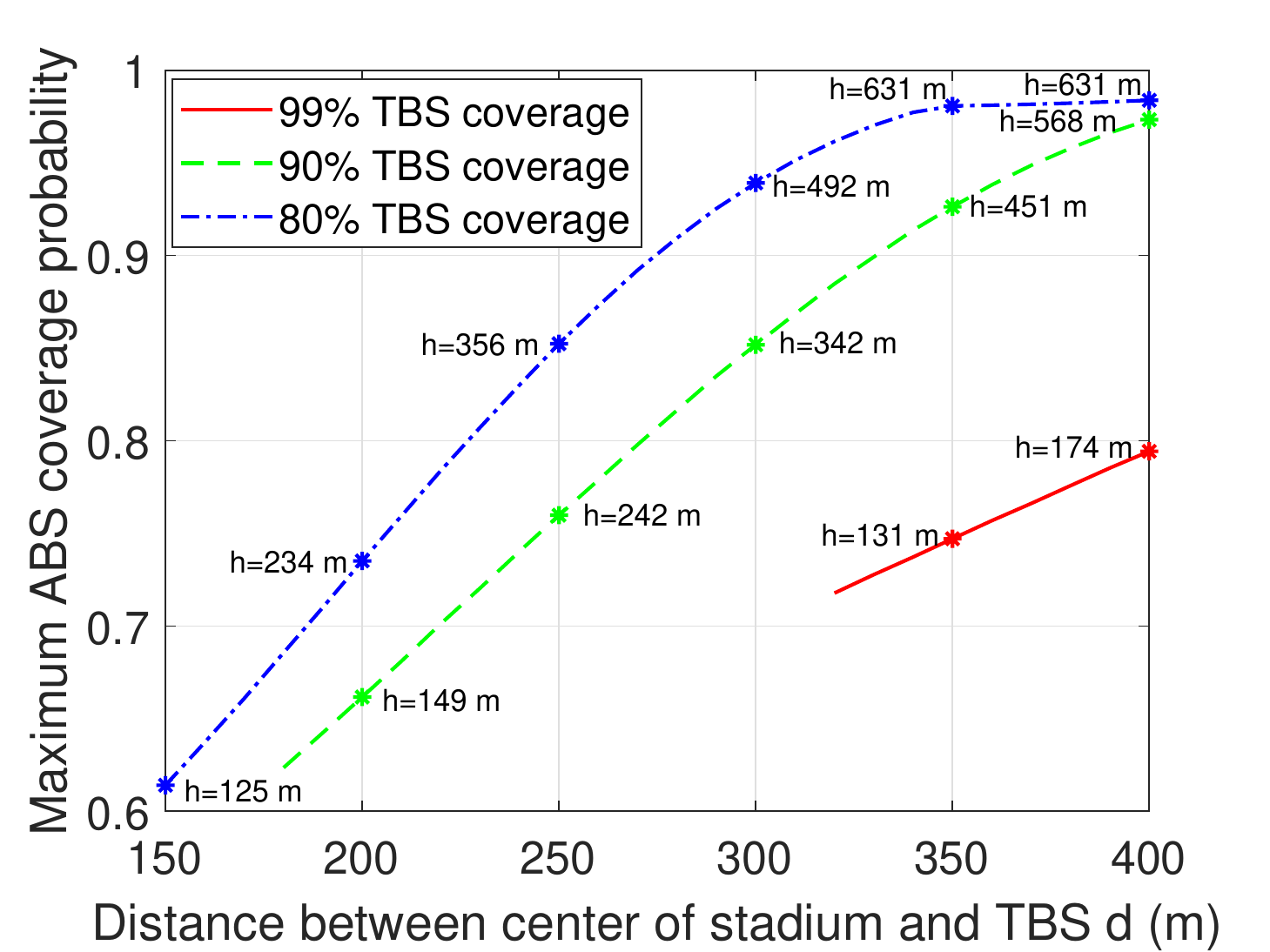}
\centering
\vspace{-2mm}
\caption{Maximum coverage probability of the ABS versus the distance between the center of the stadium and the TBS $d$ with different TBS coverage requirements.}
\centering
\label{fig:d}
\end{minipage}
\vspace{-6mm}
\end{figure*}
%\vspace{-10mm}

\textit{\underline{Impact of ABS height:}} Fig.~\ref{fig:h_bs} plots the coverage probability of the TBS against the height of the ABS with different $d$ (i.e., distance between the center of the stadium and the TBS) and different $\Pmax$ (i.e., maximum transmit power for the AsUE). From the figure, we can see that the coverage probability of the TBS first decreases as the ABS height increase. This is because the transmit power of the AsUE increases with the ABS's height, whereby the interference at the TBS increases. After a certain ABS height, the coverage probability of the TBS stays as a constant. This is due to the fact that the AsUE has reached its maximum transmit power and the interference generated at the TBS keeps the same on average. Note that the height where the coverage probability of the TBS starts to level off is independent of the projection distance between the center of the stadium and the TBS, but the height increases with the AsUE maximum transmit power.

Fig.~\ref{fig:h_uav} plots the coverage probability of the ABS against the height of the ABS with different maximum transmit power of the AsUE. We can see that the simulation results using the ATG channel model for the aerial links match very well with the analytical results\footnote{The analytical performance with the ATG channel model in~\cite{Mozaffari-2016c} is the subject of our ongoing work.}. The figure shows that there is an optimal height which can maximize the coverage probability of the ABS. This optimal height increases with the maximum transmit power of the AsUE $\Pmax$. From the figure, we find that the curves with different $\Pmax$ are overlapped when $h$ is small. This comes from the fact that in this certain range, the transmit power of the AsUE is below $\Pmax$ and the AsUE to ABS link is under full channel inversion. Under full channel inversion, the average desired signal power received at the ABS stays the same and the interference power drops as the height increases. Therefore, the coverage probability increases. When the ABS height is above the optimal height, the AsUE transmits with its maximum power $\Pmax$ and the received power of the desired signal reduces as the height increases further. Hence, the coverage probability starts to drop (e.g., the red and green curves in Fig.~\ref{fig:h_uav}). Furthermore, a larger AsUE maximum transmit power generally results in a higher coverage probability of the ABS when the height increases.

\textit{\underline{Feasibility:}} From the previous figures, we can see that there is a trade-off between the coverage probability of the TBS and the ABS. Fig.~\ref{fig:d} plots the maximum coverage probability that can be achieved at the ABS versus the distance between the center of the stadium and the TBS $d$, if $99\%$, $90\%$ or $80\%$ coverage probability of the TBS is required. The maximum achievable ABS coverage probability increases with $d$ and then becomes almost flat after a certain point.The optimal height of the ABS cannot be achieved, when $d$ is small. When $d$ is sufficiently large, the ABS can be placed at its optimal height, so the optimal coverage probability of the ABS can be achieved. For instance, from Fig.~\ref{fig:d}, we can find that for $90\%$ coverage probability of the TBS, $85\%$ coverage probability of the ABS can be achieved by placing the ABS at a height of $342$ m if the center of the stadium is $300$ m apart from the TBS. Therefore, ABS underlay cellular network is feasible under practical network setup.

\vspace{-2mm}
\section{Conclusions}\label{sec:Conc}
\vspace{-1.3mm}
In this paper, the uplink communication in a two-cell network with a TBS and an ABS was considered. We derived the exact expressions for uplink coverage probability of the TBS and the ABS in terms of the Laplace transforms of the interference power distribution at the TBS and the ABS and the distance distribution between the ABS and an i.u.d. AsUE and between the ABS and an i.u.d. TsUE. Our results demonstrated the feasibility of establishing an underlay ABS to provide uplink coverage for temporary events. Future work can consider the impact of beamforming at the ABS and user scheduling if there are multiple UEs per channel.

% The results showed that there was a trade-off between the coverage probability of the BS and the drone. Our results also showed that there was an optimal height of the drone which maximizes the coverage probability of the drone.
%
%and imperfect wireless backhaul on the coverage performance
\appendix
\setcounter{secnumdepth}{0}
\section{Appendix}
\begin{enumerate}[wide,labelwidth=!,labelindent=0pt]
	\item \textit{\underline{Proof of Lemma~\ref{le:LIB}:}}\label{ap:LIB}
Following the definition of the Laplace transform, the Laplace transform of the interference power distribution at the TBS is expressed as
\begin{small}
\begin{align}
\mathcal{L}_{\Ib}(s)&\!=\!\mathbb{E}_{\Ib}[\exp(\!-s\Ib\!)]\!=\!\mathbb{E}_{P_a,h,d}[\exp(\!-sP_a H_A d_A^{-\alphab}\!)]\nonumber\\
&=\!\mathbb{E}_{P_a,d}\left[\frac{1}{1+sP_a d_A^{-\alphab}}\right],
\end{align}
\end{small}\noindent
where the third step comes from the fact that $H_A$ follows exponential distribution with unit mean. Conditioned on the value of $h$, there are three possible cases for $\mathcal{L}_{\Ib}(s)$. When $\sqrt{(\frac{\Pmax}{\rhod})^{\frac{2}{\alphadd}}-r_2^2}<h<(\frac{\Pmax}{\rhod})^{\frac{1}{\alphadd}}$, the Laplace transform of the interference power distribution at the TBS equals to
\begin{small}
\begin{align*}
&\mathcal{L}_{\Ib}(s)\!=\!\int_h^{(\frac{\Pmax}{\rhod})^\frac{1}{\alphadd}}\!\!\mathbb{E}_{d}\left[\frac{1}{1\!+\!s\rho_Dz^{\alphadd}d_A^{-\alphab}}\right]f_{\Zd}(z)\textup{d}z\\
&+\int_{(\frac{\Pmax}{\rhod})^\frac{1}{\alphadd}}^{\sqrt{h^2+r_2^2}}\mathbb{E}_{d}\left[\frac{1}{1+s\Pmax d_A^{-\alphab}}\right]f_{\Zd}(z)\textup{d}z
\end{align*}
\end{small}
\begin{small}
\begin{subequations}
\begin{align}
&=\!\int_h^{(\frac{\Pmax}{\rhod})^\frac{1}{\alphadd}}\!\!\!\!\!\mathbb{E}_{\theta}\left[\frac{1}{1\!+\!\frac{s\rho_Dz^{\alphadd}}{\left(z^2\!-\!h^2\!+\!d^2\!-\!2\sqrt{z^2\!-\!h^2}d\cos\Theta\right)^{\frac{\alphab}{2}}}}\right]\!f_{\Zd}\!(z)\textup{d}z\nonumber\\
&+\!\int_{(\frac{\Pmax}{\rhod})^\frac{1}{\alphadd}}^{\sqrt{h^2\!+\!r_2^2}}\!\!\!\!\!\mathbb{E}_{\theta}\left[\frac{1}{1\!+\!\frac{s\Pmax}{\left(z^2\!-\!h^2\!+\!d^2\!-\!2\sqrt{z^2\!-\!h^2}d\cos\Theta\right)^{-\frac{\alphab}{2}}}}\right]\!f_{\Zd}\!(z)\textup{d}z\label{eq:prLIB1}\\
&=\int_h^{(\frac{\Pmax}{\rhod})^\frac{1}{\alphadd}}\int_{0}^{2\pi}\mathcal{L}_{\Ib}(s,\rhod z^{\alphadd}|\theta,z)\frac{1}{2\pi}f_{\Zd}(z)\textup{d}\theta \textup{d}z\nonumber\\
&+\!\!\!\int_{(\frac{\Pmax}{\rhod})^\frac{1}{\alphadd}}^{\sqrt{h^2+r_2^2}}\!\int_{0}^{2\pi}\!\!\!\mathcal{L}_{\Ib}\!(s,\!\Pmax|\theta,z)\frac{1}{2\pi}f_{\Zd}\!(z)\textup{d}\theta \textup{d}z\label{eq:prLIB2},
\end{align}
\end{subequations}
\end{small}\noindent
where $d_A$ is expressed in terms of $z$, $h$, $d$, and $\theta$ by cosine rule in \eqref{eq:prLIB1} and \eqref{eq:prLIB2} is obtained by taking the expectation over $\Theta$, which has a conditional pdf as $f_\Theta(\theta|z)=\frac{1}{2\pi}$ for $h\leqslant z\leqslant\sqrt{r_2^2+h^2}$.

Following similar steps, we can work out the Laplace transform of the interference power distribution at the TBS for the other two cases, i.e., when $h\geqslant(\frac{\Pmax}{\rhod})^{\frac{1}{\alphadd}}$ and $h\leqslant\sqrt{(\frac{\Pmax}{\rhod})^{\frac{2}{\alphadd}}-r_2^2}$.

\vspace{2.5mm}
\item \textit{\underline{Proof of Lemma~\ref{le:disUU}:}}\label{ap:disUU}
The relation between the length of the AsUE to ABS link $\Zd$ with its projection distance on the ground $r_A$ is $\Zd=\sqrt{r_A^2+h^2}$. The distance distribution of the projection distance on the ground is $f_{r_A}(r)=\frac{2r}{r_2^2}$~\cite{Guo-2014a}. Thus, we can get the pdf of $\Zd$ as
\begin{small}
\begin{align}
f_{\Zd}(z)&=\frac{\textup{d}(\sqrt{z^2-h^2})}{\textup{d}z}f_{r_A}\left(\sqrt{z^2-h^2}\right)\nonumber\\
&=\frac{z}{\sqrt{z^2-h^2}}\frac{2\sqrt{z^2-h^2}}{r_2^2}=\frac{2z}{r_2^2}.
\end{align}
\end{small}

\item \textit{\underline{Proof of Theorem~\ref{th:covBS}:}}\label{ap:covBS}
The coverage probability of the TBS is given by
\begin{small}
\begin{align}
&\Pc^\mathrm{T}\!=\!\Pr(\textsf{SINR}_\mathrm{T}>\gammab)\nonumber\\
&\!=\!\Pr\left(H_T>\frac{\gammab}{\rhob}(P_a H_A d_A^{-\alphab}+\sigma^2)\right)\nonumber\\
&\!=\mathbb{E}\!\left[\!\exp\!\left(\!-\frac{\gammab}{\rhob}(\Ib\!+\!\sigma^2)\!\right)\!\right]\!=\!\exp\!\left(\!-\frac{\gammab}{\rhob}\sigma^2\!\right)\mathcal{L}_{\Ib}(s),
\end{align}
\end{small}\noindent
where in the third step we use the fact that the link between the TsUE and the TBS experiences Rayleigh fading with a pdf of $f_{H_A}(h)=\exp(-h)$.

\vspace{2.5mm}
\item \textit{\underline{Proof of Lemma~\ref{le:disCU}:}}\label{ap:disCU}
From the proof of Lemma~\ref{le:disUU}, we know that in order to find the distribution of the distance $\Zc$ between the TsUE and the ABS, the distribution of its projection distance $r_T$ is needed. Using similar approach in~\cite{Guo-2014a,Khalid-2013}, the distance distribution of $r_T$ is derived as
\begin{small}
\begin{align}\label{eq:rc}
f_{r_T}\!(r)\!\!=\!\begin{cases}\!\!\frac{2r}{r_1^2-r_2^2},&r_2\!\leqslant \! r\!\leqslant\! r_1\!-\! d\\
            \!\!\frac{2r}{\pi r_1^2-\pi r_2^2}\mathrm{arcsec}\left(\frac{2dr}{d^2+r^2-r_1^2}\right),&r_1\!-\! d\!\leqslant \! r\!\leqslant\! r_1\!+\! d
\end{cases}.
\end{align}
\end{small}

Using the pdf of the auxiliary random variable $r_T$ in~\eqref{eq:rc}, we can obtain Lemma~\ref{le:disCU}.

\vspace{2.5mm}
\item \textit{\underline{Proof of Theorem~\ref{th:covD}:}}\label{ap:covD}
The coverage probability of the ABS is given by
\begin{small}
\begin{align}
&\Pc^\mathrm{A}\!=\!\Pr(\textsf{SINR}_\mathrm{A}>\gammad)\nonumber\\
&\!=\!\mathbb{E}_{\Zd}\left[\Pr\left(G_A>\frac{\gammad}{P_a\Zd^{-\alphadd}}(P_t G_T \Zc^{-\alphacd}+\sigma^2)|z\right)\right]\nonumber\\
&\!=\!\mathbb{E}_{\Zd}\!\!\left[\sum_{n=0}^{\mdd-1}\!\!\!\frac{(-s)^n}{n!}\exp(\!-s\sigma^2)\sum_{k=0}^n\binom{n}{k}(-\sigma^2)^{n-k}\frac{\textup{d}^k}{\textup{d}s^k}\mathcal{L}_{\Id}(s)\right]\nonumber\\
&\!=\!\int_h^{\sqrt{h^2+r_2^2}}\Pc^\mathrm{A}(P_a|z)f_{\Zd}(z)\textup{d}z,
\end{align}
\end{small}\noindent
where the third step comes from the fact that $G_A$ follows Gamma distribution. The transmit power of the AsUE $P_a$ is given in \eqref{eq:Pu}.

\end{enumerate}

\bibliographystyle{IEEEtran}

\end{document}